\documentclass[twocolumn,prl,english]{revtex4}
\usepackage{graphicx}
\usepackage{amssymb}
\usepackage{graphics}
\usepackage{times}
\usepackage{amstext}
\usepackage{bbm}
\usepackage{babel}

\begin{document}

\title{Local controllability of quantum networks}

\author{Daniel Burgarth$^{1}$}

\author{Sougato Bose$^{2}$}

\author{Christoph  Bruder$^{3}$}

\author{Vittorio Giovannetti$^{4}$}

\affiliation{$^{1}$Mathematical Institute, University of Oxford, 
24-29 St Giles' Oxford OX1 3LB, UK \\
$^{2}$Department of Physics and Astronomy, University College London, 
Gower Street,  London WC1E 6BT, UK \\
$^{3}$ Department of Physics, University of Basel, Klingelbergstrasse
82, 4056 Basel, Switzerland \\
$^{4}$NEST CNR-INFM \& Scuola Normale Superiore, Piazza dei Cavalieri
7, I-56126 Pisa, Italy}

\begin{abstract}
We give a sufficient criterion that guarantees that a many-body
quantum system can be controlled by properly manipulating the (local)
Hamiltonian of one of its subsystems.  The method can be applied to a
wide range of systems: it does not depend on the details of the
couplings but only on their associated topology.  As a special case,
we prove that Heisenberg and AKLT chains can be controlled by
operating on one of the spins at their ends. In principle, arbitrary
quantum algorithms can be performed on such chains by acting on a
single qubit.

\end{abstract}
\maketitle
\begin{figure}
\includegraphics[width=8cm]{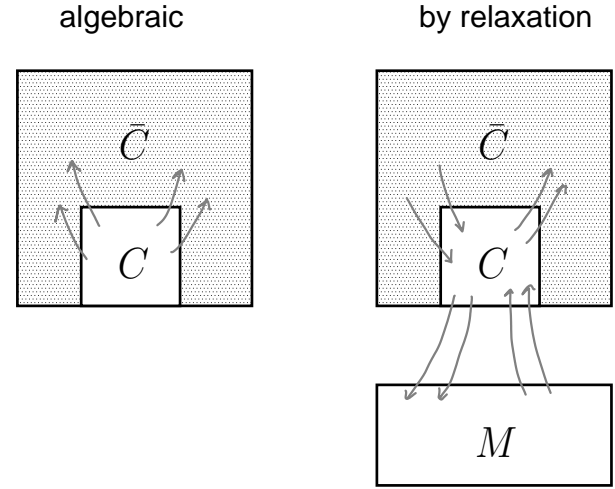}
\caption{\label{fig:schemes} 
Schematic comparison of algebraic control and control by
relaxation. 
Left panel: Controlling a subsystem $C$ of a larger
system $V= C\bigcup \overline{C}$ is sufficient to control the whole system
(algebraic control).
Right panel: The control is performed on a controlled
memory.
States on
$\overline{C}$ can be transferred from/to $M$ {\em through} $C$.}
\end{figure}

 The main obstacle in developing an efficient quantum
information technology is posed by the difficulties one faces in
achieving coherent control of quantum mechanical systems, i.e. in
externally manipulating them while preserving their quantum coherence.
There are three aspects that make control hard: firstly, quantum
systems are often rather small, so local addressing is
difficult. Secondly, control turns the quantum computer into an open
system, which introduces noise. Thirdly, due to shielding and off
resonance problems, in general there are ``invisible'' components of
an extended controlled system (say the qubits of a quantum computer)
that one cannot directly address.  It is well known that quantum
control can be simplified by properly exploiting the free Hamiltonian
evolution of the controlled
system~\cite{kane1998,Lloyd2004,Albertini2002,Ramakrishna95,Schirmer2001}.
Using this idea the problem of achieving ``complete control
everywhere'' on an extended quantum system can be reduced to ``{\em
  some} control everywhere''. In this approach {\em each} component of
the extended system is
individually~\cite{kane1998,Albertini2002,Schirmer2001,Schulte} or
jointly~\cite{Fitzsimons,Benjamin,Raussendorf} addressed by the
controlling setup, but the latter is assumed to perform only a {\em
  limited} set of allowed transformations.
While this partially solves the problem of
local addressing (at least from a theoretical perspective) and reduces
some ``harder'' quantum transformations to easier ones (for example, 2-qubit gates to
1-qubit gates), the problem of coupling the quantum system to the
external world and the problem of invisible qubits remains. 
One way to cope with  these issues, is to replace the ``some control
everywhere'' approach with a ``complete control \emph{somewhere}'' approach, where ``somewhere''
is ideally a small portion $C$ of a larger system
$V=C\bigcup\overline{C}$  that we want to control ($\overline{C}$ being the part of $V$ on which we do not have direct access).
In this scenario two alternative control techniques have been proposed so far:
an {\em algebraic control} (AC) method~\cite{Lloyd2004,Romano2006,kay2008,Schirmer2008}
and a {\em control by relaxation} (RC) method~\cite{Burgarth2007,Burgarth2007c} (see Fig.~\ref{fig:schemes}).
In the former case one assumes unlimited direct access on $C$ 
by means of time-dependent local Hamiltonians which are properly modulated ---  see below for details.  
In the latter case 
instead one assumes to operate on $C$  by means of a limited set of 
quantum gates that couple it with some external, completely controlled, quantum memory $M$.
Here the control
is realized by transferring the states of $V$
into $M$ where their are manipulated (e.g. transformed or measured)
and then transferred back to
$V$.

In this article 
we show that 
an easy-to-check graph infection criterion that has been developed for control by relaxation~\cite{Burgarth2007}
can also be used for algebraic control. This is a major improvement with respect to previous  works on the subject~\cite{Lloyd2004,Romano2006,kay2008,Schirmer2008} since it allows us to check AC controllability
 of {\em large} many-body systems. 
As a special case we prove
that Heisenberg spin chains of {\em arbitrary length} admit
algebraic control when acted upon at one end spin only. This gives the first non-trivial
example of such mediated control and has important consequences for quantum computation:
 in principle, arbitrary quantum algorithms can be performed on such chains by acting on a single qubit. 
The controllability even holds when
magnetic fields spoil the conservation of excitations (in this case the criterion is no
longer applicable to RC). 
Since the criterion developed here is
of topological nature, it does not depend on the details of the
couplings and can therefore be applied to a wide range of experimental
realizations of many-body quantum information processing, ranging from
optical lattices~\cite{optical}, to arrays of coupled
cavities~\cite{arrays,arraysb}, to solid state
qubits~\cite{kane1998,solid}.
Finally, the issue of how local
time dependent terms in the Hamiltonian can influence the global dynamics is also
interesting from a purely theoretical perspective.

\paragraph*{Algebraic control: --} 
We start by reviewing the basic properties of AC.
In algebraic control~\cite{Lloyd2004,Albertini2002,Ramakrishna95,Schirmer2001}
 the composite system $V = C\bigcup \overline{C}$ is described by
a global Hamiltonian of the form
  $H_V+\sum_{k}f_{k}(t)\;
  h_{C}^{(k)}\otimes\openone_{\overline{C}}$.
Here $H_V$ is some fixed coupling Hamiltonian on $V$ while $h_{C}^{(k)}$ are
a set of local controlling Hamiltonians operating on $C$ that can be activated through the (time dependent) modulating parameters $f_k(t)$.
At the mathematical level, a general necessary and sufficient criterion 
for this scheme has been derived~\cite{Lloyd2004,Albertini2002,Ramakrishna95,Schirmer2001}. It states that $V$ 
is AC controllable by properly tuning the functions  $f_k(t)$
 iff $iH_V$ and $i h_{C}^{(k)}$ are {\em generators} of the
Lie algebra ${\cal L}(V)$ of the composite system $V$
(the set of all skew-Hermitian operators of $V$), i.e. 
\begin{equation} \left\langle i H_V,\mathcal{L}(C)\right\rangle
=\mathcal{L}(V),\label{eq:imp02}
\end{equation} 
where, for the sake of simplicity, we have assumed the $i h_{C}^{(k)}$'s
to be generators of the local Lie algebra ${\cal L}(C)$ of $C$ and where 
we use the symbol $\left\langle
\mathcal{A},\mathcal{B}\right\rangle $ to represent the algebraic closure of
the operator sets $\mathcal{A}$ and $\mathcal{B}$.
In simpler terms this implies that 
 any possible quantum transformation on $V$ can be operated
by acting on $C$ iff 
 all elements of ${\cal L}(V)$ can be obtained
as a linear combinations of $i H_V$, $\mathcal{L}(C)$ and
iterated commutators of these operators.

Although the general arguments in~\cite{Lloyd2004} suggest that \emph{most} quantum systems satisfy 
the criterion~(\ref{eq:imp02}), up to now only few examples 
have been presented~\cite{Romano2006,kay2008,Schirmer2008}. 
Indeed the condition~(\ref{eq:imp02})
 can be tested numerically only for relatively small systems (say maximally ten qubits). It becomes impractical instead  when applied  to large many-body systems  
where $V$ is a collection of  quantum sites (e.g. spins) whose Hamiltonian
is described as a summation of two-sites terms.   
The main result of
this paper is 
the derivation of an {\em inductive} easy-to-check method 
to test the AC controllability condition~(\ref{eq:imp02}) for such configurations.

\paragraph*{Graph Criterion:--}
The proposed method exploits
the topological properties of the graph defined by the coupling terms entering the
 many-body Hamiltonian $H_{V}$. 
This allows us to translate the AC controllability problem into a simple graph infection  property 
which can be  easily tested. 
We start reviewing the latter 
for the most general setup, 
which will show more
clearly where the topological properties come from. 

\begin{figure}
\includegraphics[width=8cm]{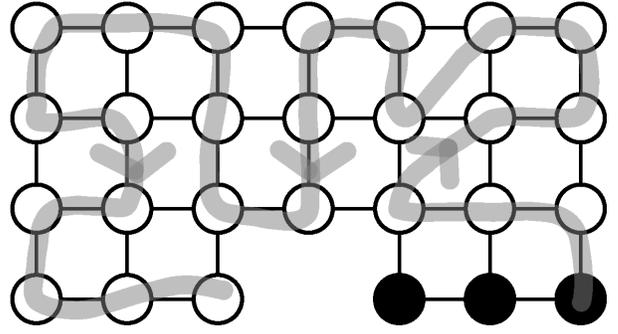}
\caption{\label{fig:order} Example for a graph that fulfills the
  infection criterion. The set $C$ of ``infected nodes'' is given by
  the black nodes, and the grey line shows the order in which new
  nodes can be added to the set $P_k$ such that
  Eqs.~(\ref{eq:oneneigh},\ref{eq:unique}) hold.}
\end{figure}

 The graph
infection property was introduced in~\cite{Burgarth2007} and analyzed
from a purely graph theoretical perspective
in~\cite{Aazami2008}.  In words, the infection process can be described
as follows: an initial set of nodes of the graph is ``infected''. The infection then
spreads by the following rule: an infected node infects a ``healthy'' neighbor
if and only if it is its \emph{only} healthy neighbor. If eventually
all nodes are infected, the initial set is called \emph{infecting}.
More formally, we consider an undirected graph
$G=(V,E)$ characterized by a set of nodes $V$ and by a set of edges
$E$, and a subset $C\subseteq V$. 
We call $C$ \emph{infecting} $G$ iff there exists an ordered sequence
$\{ P_{k}; k=1, 2, \cdots , K\}$ of $K$ subsets of
$V$ 
\begin{equation} C=P_{1}\subseteq P_{2}\subseteq\cdots\subseteq
  P_{k}\subseteq\cdots\subseteq P_{K}=V\;,\label{eq:sequence}
\end{equation} 
such that each set is exactly one node larger than the previous
one,
\begin{equation} P_{k+1}\backslash P_{k}=\left\{m_{k}\right\},  
\label{eq:oneneigh}
\end{equation} 
and there exists an $n_k \in P_{k}$ such that $m_{k}$ is its unique
neighbor \emph{outside} $P_{k}:$ 
\begin{equation} N_{G}(n_{k})\cap V\backslash P_{k}=\left\{m_{k}\right\},
\label{eq:unique}
\end{equation} 
with $N_G(n_k) \equiv \{ n \in V| (n,n_k) \in E \}$ being the set of
nodes of $V$ which are connected to $n_k$ through an element of $E$.
The sequence $P_{k}$ provides a natural structure
(Fig.~\ref{fig:order}) on the graph which allows us to treat it almost
as a chain (although the graphs can be very much different from
chains, see also the examples given in~\cite{Burgarth2007}).
In particular, it gives us an index $k$ over which we will be able
to perform inductive proofs.

 The link to quantum mechanics is that
each node $n$ of the graph has a quantum degree of freedom associated with the Hilbert space
 $\mathcal{H}_{n}$, which describes the $n$-th site  of the many-body system $V$ we wish to control.
The coupling Hamiltonian determines the edges through
\begin{equation}
H_{V}=\sum_{(n,m)\in E}H_{nm}\;,\label{eq:ham}
\end{equation}
where $H_{nm}=H_{mn}$ are some arbitrary Hermitian operators acting on
$\mathcal{H}_{n}\otimes\mathcal{H}_{m}$.
Within this  context we call the
Hamiltonian~(\ref{eq:ham}) \emph{algebraically propagating} iff for
all $n \in V$ and $(n,m)\in E$ one has,
\begin{equation} \left\langle \left[ i
    H_{nm},\mathcal{L}(n)\right],\mathcal{L}(n)\right\rangle
  =\mathcal{L}(n,m),\label{eq:important}
\end{equation} 
where for a  generic set of nodes $P\subseteq V$, 
${\cal L}(P)$ is the Lie algebra  associated with the Hilbert
space $\bigotimes_{n\in P}\mathcal{H}_{n}$~\cite{NOTA11}.
The graph criterion can then be expressed as follows:

\begin{description}
\item [{Theorem:}] 
\emph{Assume that the Hamiltonian~(\ref{eq:ham}) of the composed system $V$ is algebraically propagating
and that $C\subseteq V$ infects $V$. Then $V$ is algebraically controllable acting on its subset $C$.} 
 \item [{Proof:}]
To prove the theorem we have to show that Eq.~(\ref{eq:imp02}) holds, or 
equivalently that $\mathcal{L}(V)\subseteq\left\langle i H_{V},\mathcal{L}(C)\right\rangle$
(the opposite inclusion  being always verified). 
 To do so we proceed by induction over $k=1,\ldots,K$, showing that 
$\mathcal{L}(P_{k})\subseteq\left\langle i H_{V},\mathcal{L}(C)\right\rangle$. 
\end{description}
Basis: by Eq.~(\ref{eq:sequence}) we have 
$\mathcal{L}(P_{1})=\mathcal{L}(C)\subseteq\left\langle i
H_{V},\mathcal{L}(C)\right\rangle$ .
Inductive step: assume that for some $k<K$
\begin{equation}
\mathcal{L}(P_{k})\subseteq\left\langle iH_{V},\mathcal{L}(C)\right\rangle .
\label{eq:ind}
\end{equation}
We now consider $n_{k}$ from Eq.~(\ref{eq:unique}). We have 
$\mathcal{L}(n_{k})\subset\mathcal{L}(P_{k})
\subseteq\left\langle i H_{V},\mathcal{L}(C)\right\rangle$
and 
\begin{eqnarray}\nonumber
\left[i H_{n_{k},m_{k}},\mathcal{L}(n_{k})\right]=\left[iH_{V},\mathcal{L}
(n_{k})\right]-\sum_{m}\left[i H_{n_{k},m},\mathcal{L}(n_{k})\right],
\end{eqnarray}
where the sum on the right hand side contains only nodes from
$P_{k}$ by Eq.~(\ref{eq:unique}). It is therefore an element of
$\mathcal{L}(P_{k})$. The first term on the right hand side is a
commutator of an element of $\mathcal{L}(P_{k})$ and $i H_{V}$ and thus an
element of $\left\langle i H_{V},\mathcal{L}(C)\right\rangle $ by
Eq.~(\ref{eq:ind}).  Therefore 
$\left[ i H_{n_{k},m_{k}},\mathcal{L}(n_{k})\right]\subseteq
\left\langle i H_{V},\mathcal{L}(C)\right\rangle $ and by algebraic propagation
Eq.~(\ref{eq:important}) we have 
\begin{eqnarray} \left\langle \left[i
    H_{n_{k},m_{k}},\mathcal{L}(n_{k})\right],\mathcal{L}(n_{k})\right\rangle
  =\mathcal{L}(n_{k},m_{k})\subseteq\left\langle i
  H_{V},\mathcal{L}(C)\right\rangle .\nonumber
\end{eqnarray} 
But $\left\langle \mathcal{L}(P_{k}),\mathcal{L}(n_{k},m_{k})\right\rangle
=\mathcal{L}(P_{k+1})$ by Eq.~(\ref{eq:oneneigh}) so 
$  \mathcal{L}(P_{k+1})\subseteq\left\langle i
  H_{V},\mathcal{L}(C)\right\rangle $.
Thus by induction 
\begin{equation}
\mathcal{L}(P_{K})=\mathcal{L}(V)\subseteq\left\langle i
  H_{V},\mathcal{L}(C)\right\rangle
  \subseteq\mathcal{L}(V).\qquad\blacksquare
\newline
\end{equation}

The above theorem has split the question of algebraic control into two
separate aspects. The first part, the algebraic propagation
Eq.~(\ref{eq:important}) is a property of the coupling that lives on a
small Hilbert space $\mathcal{H}_{n}\otimes\mathcal{H}_{m}$ and can
therefore be checked easily numerically -- we have for instance
verified this property for Heisenberg-like (see below),
AKLT~\cite{AKLT}, and for SU(3) Hamiltonians~\cite{SU3}.  The second
part is a topological property of the (classical) graph. An important
question arises here if this may be not only a sufficient but also
necessary criterion. As we will see below, there are systems where $C$
does not infect $V$ but the system is controllable \emph{for specific
  coupling strengths}. However the topological stability with respect
to the choice of coupling strengths is no longer given.

\paragraph*{Application to spin networks:--}

An important example of the above theorem are systems of coupled
spin-$1/2$ systems (qubits). We consider the two-body Hamiltonian
given by the following Heisenberg-like
coupling,
\begin{equation} \label{equat1}
H_{nm}=c_{nm}\left(X_{n}X_{m}+Y_{n}Y_{m}+\Delta Z_{n}Z_{m}\right)\;,
\end{equation}
where the $c_{nm}$ are arbitrary coupling constants, $\Delta$ is an
anisotropy parameter, and $X$, $Y$, $Z$ are the standard Pauli
matrices. The edges of the graph are those $(n,m)$ for which
$c_{nm}\neq0$.  The relaxation controllability of this model was
extensively analyzed in Refs.~\cite{Burgarth2007,Burgarth2007c} while,
in the restricted case of the single excitation subspace, its
algebraic controllability was exactly solved in
Ref.~\cite{Schirmer2008,BGB2007}.  

\begin{figure}
\includegraphics[width=8cm]{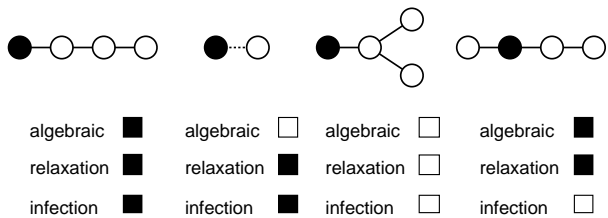}
\caption{\label{fig:Examples-for-small}Examples
of small spin networks and their controllability. The black circles
are the controlled parts $C$ and the white ones the uncontrolled
part $\overline{C}$. The dashed line stands for a coupling of the
$X_{n}X_{m}+Y_{n}Y_{m}$ type (which is not algebraically
propagating), and the solid line stands for the full (possibly anisotropic)
Heisenberg coupling~(\ref{equat1}) with $\Delta\neq 0$.
The filled/empty boxes below each graph indicate whether or not 
it can be controlled by algebraic methods, by relaxation, and 
whether the graph infection property holds.
For the last two examples the graph is not infected, and
controllability depends explicitly on the coupling strengths (all of
which are assumed to be equal).}
\end{figure}

To apply our method we have first shown that the Heisenberg
interaction is algebraically propagating.  In this case the Lie
algebra ${\cal L}(n)$ is associated to the group $\mbox{su}(2)$ and it
is generated by the operators $\{ i X_n, i Y_n, i Z_n\}$.  Similarly
the algebra ${\cal L}(n,m)$ is associated with $\mbox{su}(4)$ and it
is generated by the operators $\{ i X_n I_m , i X_n X_m,i X_n Y_m,
\cdots , i Z_nZ_m\}$.  The identity (\ref{eq:important}) can thus be
verified by observing that
\begin{eqnarray*}
\left[X_{n},H_{nm}\right] & = & Z_{n}Y_{m}-Y_{n}Z_{m}\\
\left[Z_{n},Z_{n}Y_{m}-Y_{n}Z_{m}\right] & = & X_{n}Z_{m}\\
\left[Y_{n},X_{n}Z_{m}\right] & = & Z_{n}Z_{m}\\
\left[X_{n},Z_{n}Z_{m}\right] & = & Y_{n}Z_{m},\end{eqnarray*}
where for the sake of simplicity irrelevant constants have been removed. 
Similarly using the cyclicity $X\rightarrow Y\rightarrow Z\rightarrow X$ of
the Pauli matrices we get, \begin{eqnarray*}
X_{n}Z_{m} & \rightarrow & Y_{n}X_{m}\rightarrow Z_{n}Y_{m}\\
Z_{n}Z_{m} & \rightarrow & X_{n}X_{m}\rightarrow Y_{n}Y_{m}\\
Y_{n}Z_{m} & \rightarrow & Z_{n}X_{m}\rightarrow X_{n}Y_{m}.\end{eqnarray*}
Finally, using \[
\left[Z_{n}Z_{m},Z_{n}Y_{m}\right]=X_{m}\;,\]
and cyclicity, we obtain all $15$ basis elements of ${\cal L}(n,m)$ concluding the proof. 
According to our Theorem we can thus conclude that {\em any} network of 
spins coupled through  Heisenberg-like interaction 
 is  AC controllable when
operating on the subset $C$, if 
the associated graph can be infected. 
In particular, this shows that Heisenberg-like chains with arbitrary coupling strengths admits AC controllability 
 when operated at one end (or, borrowing from~\cite{Lloyd2004}, that 
the extreme of such chains are universal quantum interfaces for the whole system). 
We remark that in this case, knowledge about the coupling parameters
of the Hamiltonian can be obtained by controlling one end qubit
only \cite{Koji2008}. 
The case $\Delta=0$ on the other hand is an
interesting example where relaxation control is possible but our
theorem cannot be applied.  Using the numerical method
from~\cite{Schirmer2001} we found that already a chain of length $N=2$
cannot be controlled by acting with arbitrary Pauli operators on one
end -- see also Ref.~\cite{Romano2006}. For the case $\Delta\neq0$, a
star with $N=4$ provides a good example that
property~(\ref{eq:important}) without graph infection does not suffice
to provide controllability~(Fig.~\ref{fig:Examples-for-small}).
Another interesting example is an Ising chain with a magnetic field in 
a generic direction, which is controllable for $N=2,3$ but, 
perhaps surprisingly, not for longer chains.
Finally, we have confirmed that SU(3) and AKLT Hamiltonians are 
algebraically propagating. These interactions have the form
\begin{equation}
H_{nm}=c_{nm}\left( A (S_n \cdot S_m)^2 + B S_n \cdot S_m \right)\;,
\end{equation}
where $S_n$ is the spin operator of particle $n$. The analytical
method sketched above for the Heisenberg chain turns out to be quite
cumbersome, so we used the numerical methods given in
\cite{Schirmer2001} to check that Eq.~(\ref{eq:important})
holds. Since Eq.~(\ref{eq:important}) lives in a small Hilbert space,
this computation is efficient and fast. It then follows by our theorem
that these systems are controllable for arbitrary length.

\paragraph*{Conclusions:--}

In this paper we have presented a criterion to determine if a
many-body quantum system allows for algebraic control by operating on
a proper subset of it.  In contrast to previous proposals the method
does not require the knowledge of the spectrum of the system
Hamiltonian.  Instead it exploits some topological properties of the
graph associated with its coupling terms.  As a special case, we have
proven that Heisenberg and AKLT chains can be controlled by operating
on one of the spins at their ends. In principle, arbitrary quantum
algorithms can be performed on such chains by acting on a single
qubit.  This motivates the search for further explicit specific and
efficient control schemes on spin chains. 

\begin{acknowledgments}
We acknowledge fruitful discussions with T. Schulte-Herbr\"uggen, 
support by the QIP-IRC and Wolfson College Oxford
(D.B.), the EPSRC, UK, the Royal Society and the Wolfson
Foundation (S.B.),  the EC IST-FET project EuroSQIP, the Swiss
NSF, and the NCCR Nanoscience (C.B.), and the Quantum Information
research program of Centro Ennio De Giorgi of SNS (V.G.).
\end{acknowledgments}


\begin{thebibliography}{10}
\bibitem{NOTA11} Note that the condition (\ref{eq:important})
is a stronger property than the condition of controlling $n,m$ by 
acting on $n$. According to Eq.~(\ref{eq:imp02})
the latter in fact reads
$\left\langle i H_{nm},\mathcal{L}(n)\right\rangle
=\mathcal{L}(n,m)$, 
which is implied by Eq.~(\ref{eq:important}).
\end{thebibliography}
\end{document}